\documentstyle[sprocl,amsmath,psfig]{article}

\begin{document}

\title{EFFECTIVE POTENTIAL APPROACH
       TO QUANTUM DISSIPATION IN CONDENSED MATTER SYSTEMS}

\author{Alessandro Cuccoli$^{a,c}$, Andrea Fubini$^{a,c}$,
        Valerio Tognetti$^{a,c}$,\\ Ruggero Vaia$^{b,c}$}

\address{$^a$~Dipartimento di Fisica dell'Universit\`a di Firenze,\\
              Largo E. Fermi~2, I-50125 Firenze, Italy. \\
         $^b$~Istituto di Elettronica Quantistica
              del Consiglio Nazionale delle Ricerche, \\
              via Panciatichi~56/30, I-50127 Firenze, Italy.\\
         $^c$~Istituto Nazionale di Fisica della Materia (INFM),
              Unit\`a di Firenze.}

\maketitle

\abstracts{
The effects of dissipation on the thermodynamic properties of nonlinear
quantum systems are approached by the path-integral method in order
to construct approximate classical-like formulas for evaluating thermal
averages of thermodynamic quantities. Explicit calculations are
presented for one-particle and many-body systems. The effects of the
dissipation mechanism on the phase diagram of two-dimensional
Josephson arrays is discussed.}

\section{Introduction}

The usefulness of the improved~\cite{GTall} effective potential
approach,~\cite{FeynmanH65} has been proven by several applications
to condensed matter systems.  However, open systems were not
immediately treated and previous studies were confined to obtain a
classical-like expression for the free energy.~\cite{Weiss99,BaoZW95}

In fact, the effective-potential method, also called {\em pure-quantum}
{\em self-con\-sistent} {\em harmonic approximation} (PQSCHA) after its
generalization to phase-space Hamiltonians,~\cite{CTVV92ham,CGTVV95}
is able~\cite{CRTV97} to give the density matrix of a nonlinear
system interacting with a dissipation bath through the
Caldeira-Leggett (CL) model.~\cite{CaldeiraL83}
For a better understanding of the method let us first consider
one single degree of freedom. In the general case
the CL model starts from the Hamiltonian
\begin{equation}
 \hat{H}={\hat {p}^2\over2m}+V(\hat {q}) +{1\over2}\sum_i\bigg\{{\hat
 {p}_i^2\over m_i} + m_i\omega_i^2\Big[\hat {q}_i-F_i(\hat {q})\Big]^2
 \bigg\}~, \label{e.HCL}
\end{equation}
where $(\hat{p},\hat{q})$ and $(\hat{p}_i,\hat{q}_i)$ are the momenta
and coordinates of the system and of an environment (or bath) of
harmonic oscillators.
When $F_i(\hat{q})=\hat{q}$ the dissipation is said to be {\em linear}
and we will restrict ourselves to this case in the following. The bath
coordinates can be integrated out exactly from the corresponding path
integral and the CL Euclidean action is obtained in the form:
\begin{equation}
 S[q(u)] =\!\int_0^{\beta} \!\!\! du \left[ {m\over2}\dot q^2(u) {+}
 V\big(q(u)\big) \right] +\! \int_0^{\beta} \! {du\over2} \!
 \int_0^{\beta}\!\!\!\! du'\,k(u{-}u')\,q(u)\,q(u')\,.
 \label{e.S}
\end{equation}
To make contact with the classical concept of dissipation, one may
take the classical counterpart of Eq.~(\ref{e.HCL}) and get the
Langevin equation of motion:
\[
 m\ddot q + m\int_0^\infty\!\! dt'\,\gamma(t')\, \dot q(t{-}t') +
 V'(q) = 0~;
\]
then the Laplace transform $\gamma(z)$ of the {\em damping function}
$\gamma(t)$, as expressed in terms of the spectral density of the
environmental coupling,~\cite{Weiss99} can be related to the Matsubara
components of the {\em damping kernel} $k(u)$ ($\nu_n=2\pi{n}/\beta$)\,:
\begin{equation}
 k(u)\equiv m\beta^{-1}~{\textstyle{\sum}_n}~e^{i\nu_nu}\,k_n~,
 ~~~~~~~~~~~ k_n=|\nu_n|\,\gamma\big(z{=}|\nu_n|\big)~,
 \label{e.kgamma}
\end{equation}
We will consider two cases:
\begin{eqnarray*}
 &{\rm Ohmic:~~~~~~} &\gamma(t)=\gamma\,\delta(t-0)~, ~~~~~~~\,
 \gamma(z)=\gamma~,
\\
 &{\rm Drude:~~~~~~}
 &\gamma(t)=\gamma\,\omega_{\rm{D}}\,e^{-\omega_{\rm{D}}t}~, ~~~~~~
 \gamma(z)={\gamma\,\omega_{\rm{D}}/(\omega_{\rm{D}}+z)}~,
\end{eqnarray*}
where the dissipation strength $\gamma$ and the bath bandwidth
$\omega_{\rm{D}}$ characterize the environmental coupling;
$\omega_{\rm{D}}\to\infty$ gives the Ohmic (or Markovian) case.
In order to interpolate between different regimes and to take
into account memory effects we can also assume:
\begin{equation}
 \gamma(z)\propto z^s~,
\label{gamma}
\end{equation}
with $-1<s<1$ and $s>0$ ($s<0$) is said the {\em super-} ({\em sub-})
Ohmic case.

\section{Effective potential in presence of dissipation \label{PQSCHA}}
\medskip

The PQSCHA approximation consists in taking as trial action $S_0$, the
most general quadratic functional with the same linear dissipation of $S$,
namely,
\begin{eqnarray}
 S_0[q(u)] &=& \int_0^{\beta} du \left[ {m\over2}\dot q^2(u) + w
 + m\omega^2\,\big(q(u){-}\bar{q}\big)^2 \right]
\nonumber\\
 & & \hspace{20mm} +\! \int_0^{\beta} \! {du\over2} \!
 \int_0^{\beta}\!\!\!\! du'\,k(u{-}u')\,q(u)\,q(u')\,.  \label{e.S0}
\end{eqnarray}
The quantity $\bar{q}=\beta^{-1}\int{q(u)}\,du$ is the average point
of paths and
\begin{equation}
 w=w(\bar{q})~,~~~~~~~
 \omega^2=\omega^2(\bar{q})~,
\end{equation}
are parameters to be determined by minimizing the right-hand side of
the {\em Feynman inequality},
\begin{equation}
F\leq{F_0}+\beta^{-1}\langle S-S_0\rangle_{S_0}~.
\end{equation}
For any observable $\hat{\cal O}(\hat{p},\hat{q})$ the $S_0$-average
can be expressed~\cite{CGTVV95,CRTV97} in terms of its Weyl
symbol~\cite{Berezin80} ${\cal O}(p,q)$. As final result we obtain the
classical-like form
\begin{equation}
 \langle\hat{{\cal O}}\rangle =\frac1{{\cal Z}_0}
 \sqrt{m\over{2\pi\hbar^2\beta}} \int\,d\bar{q}
 ~\big\langle\!\big\langle{\cal O}(p,\bar{q}+\xi)\big\rangle\!\big\rangle
 ~e^{-\beta\,V_{\rm eff}(\bar{q})}~,
\label{e.aveOpq}
\end{equation}
where $\langle\!\langle{\cdots}\rangle\!\rangle$ is a Gaussian average
operating over $p$ and $\xi$ with moments
\begin{eqnarray}
 \langle\!\langle{\xi^2}\rangle\!\rangle \equiv \alpha(\bar{q}) &=&
 {2\over\beta m}~ \sum\limits_{n=1}^{\infty}
 {1\over \nu_n^2{+}\omega^2(\bar{q}){+}k_n}
 ~~\xrightarrow[k\to{0}]~~
 {1\over2m\omega}\Big(\coth f{-}{1\over f}\Big)~,
\label{e.alpha}
\\
 \langle\!\langle{p^2}\rangle\!\rangle \equiv \lambda(\bar{q}) &=&
 {m\over\beta}\,\sum\limits_{n=-\infty}^{\infty}
 {\omega^2(\bar{q})+k_n \over \nu_n^2{+}\omega^2(\bar{q}){+}k_n}
 ~~\xrightarrow[k\to{0}]~~
 {m\omega\over2}\coth f ~,
\label{e.lambda}
\end{eqnarray}
where $f\equiv\beta\omega/2$.
The effective potential is defined as $ V_{\rm eff}(\bar{q})\equiv
w(\bar{q})+\sigma(\bar{q})$~, with
\begin{equation}
 \sigma(\bar{q})={1\over\beta}\sum_{n=1}^{\infty}
 \ln\,{\nu_n^2{+}\omega^2(\bar{q}){+}k_n \over \nu_n^2}
 ~~\xrightarrow[k\to{0}]~~
  {1\over\beta}\,\ln{\sinh f\over f} ~.
\label{e.sigma}
\end{equation}
The r.h.s. of Eqs.~(\ref{e.alpha}-\ref{e.sigma}) show that the well known
non-dissipative limits are recovered for ${k\to{0}}$. The variational
parameters can be self-consistently calculated and the explicit
expressions are the following
\begin{equation}
 w(\bar{q})=\big\langle\!\big\langle V(\bar{q}+\xi)\big\rangle\!\big\rangle
 -{\textstyle \frac 12} m\omega^2\alpha~,
\hspace{10mm}
 m\,\omega^2(\bar{q}) = \big\langle\!\big\langle V''(\bar{q}+\xi)\big\rangle\!\big\rangle~.
\end{equation}

\section{Applications}
\medskip

\begin{figure}[b!]
\centerline{\psfig{bbllx=16mm,bblly=80mm,bburx=192mm,bbury=200mm,%
figure=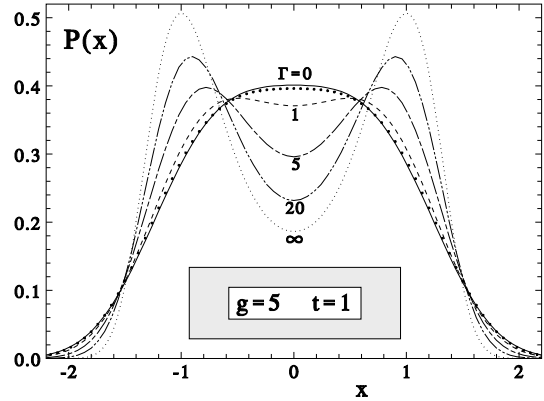,width=70mm,angle=0}}
\caption{Configuration density $P(x)=\langle\delta(\hat{x}-x)\rangle$ of
the double-well quartic potential for the fixed coupling $g=5$, the
reduced temperature $t=1$, and different values of the Ohmic damping
parameter $\Gamma=\gamma/\omega_0$, being $\omega_0$ the
characteristic frequency of the system. The filled circles are the
exact result for $\Gamma=0$; the dotted curve at $\Gamma=\infty$
corresponds to the classical limit.
\label{f.1}}
\end{figure}

In order to understand how this approximation scheme works, let us
take first a very simple system: one particle in a double-well
potential with Ohmic dissipation. A typical result for this system
is shown in Fig.~\ref{f.1}: the dissipation quenches the quantum
fluctuations of the coordinate. However, it must be noted from
Eq.(\ref{e.lambda}) that those of the momentum are infinite.

Let us now turn to the many-body case. The first application is a
quantum $\varphi^4$-chain of particles with Drude-like
dissipation:~\cite{CFTV99} the undamped system is described by the
following action
\begin{equation}
 S_{\varphi^4}={3\over{2Q}}\int_0^\beta du\,a\,\sum_i\left[{\dot
 q_i^2\over2} +{(q_i-q_{i-1})^2\over2a^2} + {\Omega^2\over8}
 \big(1-q_i^2\big)^2 \right]~,
\label{e.Sphi4}
\end{equation}
where $a$ is the chain spacing, $\Omega$ is the gap of the bare
dispersion relation, and $Q$ is the quantum coupling.  The classical
continuum model supports kink excitations of characteristic width
$\Omega^{-1}$ and static energy
$\varepsilon_{\rm{K}}=\Omega/Q$. This energy is used as the energy
scale in defining the reduced temperature
$t\equiv(\beta\varepsilon_{\rm{K}})^{-1}$. We also use the kink length
in lattice units $R\equiv(a\Omega)^{-1}$ ($R\to\infty$ in the
continuum limit). We assume uncorrelated identical CL baths for each
degree of freedom, and use the low-coupling
approximation~\cite{CGTVV95} (LCA) in order to deal with the effective
potential. The partition function turns out to be written as
\begin{eqnarray}
 {\cal{Z}}&=& ({3t\over4\pi RQ^2})^{N\over2}\!\!\!\int\! d^N \! q\;
 e^{-\beta V_{\rm eff}} ~,
\label{e.Z}\\
 \beta V_{\rm eff}&=&\frac{3}{2Rt} {\sum}_i\left[{R^2\over2}
 (q_i-q_{i-1})^2 + v_{\rm eff}(q_i) \right]~,
\label{e.Vfphi4}\\
 v_{\rm eff}(q)&=&{1\over8}\big(1-3D-q^2\big)^2 + {3\over4}D^2 +{Rt\over
 N}{\sum}_k\sigma_k~.
\label{e.vfphi4}
\end{eqnarray}

The renormalization parameter $D=D(t;Q,R;\gamma,\omega_{\rm{D}})$
generalizes Eq.~(\ref{e.alpha}) and is the solution of the
self-consistent equations:
\begin{eqnarray}
 D&=&{4Rt\over3N}{\sum}_k{\sum}_n
 {\Omega^2\over\nu_n^2+\omega_k^2+k_n},
\\
 \omega_k^2(t)&=&\Omega^2\big[(1-3D)+4R^2\sin^2(ka/2)\big]~,
\end{eqnarray}
where $k_n=\gamma\omega_{\rm{D}}\,\nu_n/(\omega_{\rm{D}}+\nu_n)$ and
here $\nu_n/\Omega=2\pi{n}t/Q$.

\begin{figure}[b!]
\begin{minipage}[t]{119mm}
\centerline{\psfig{bbllx=20mm,bblly=20mm,bburx=188mm,bbury=238mm,%
figure=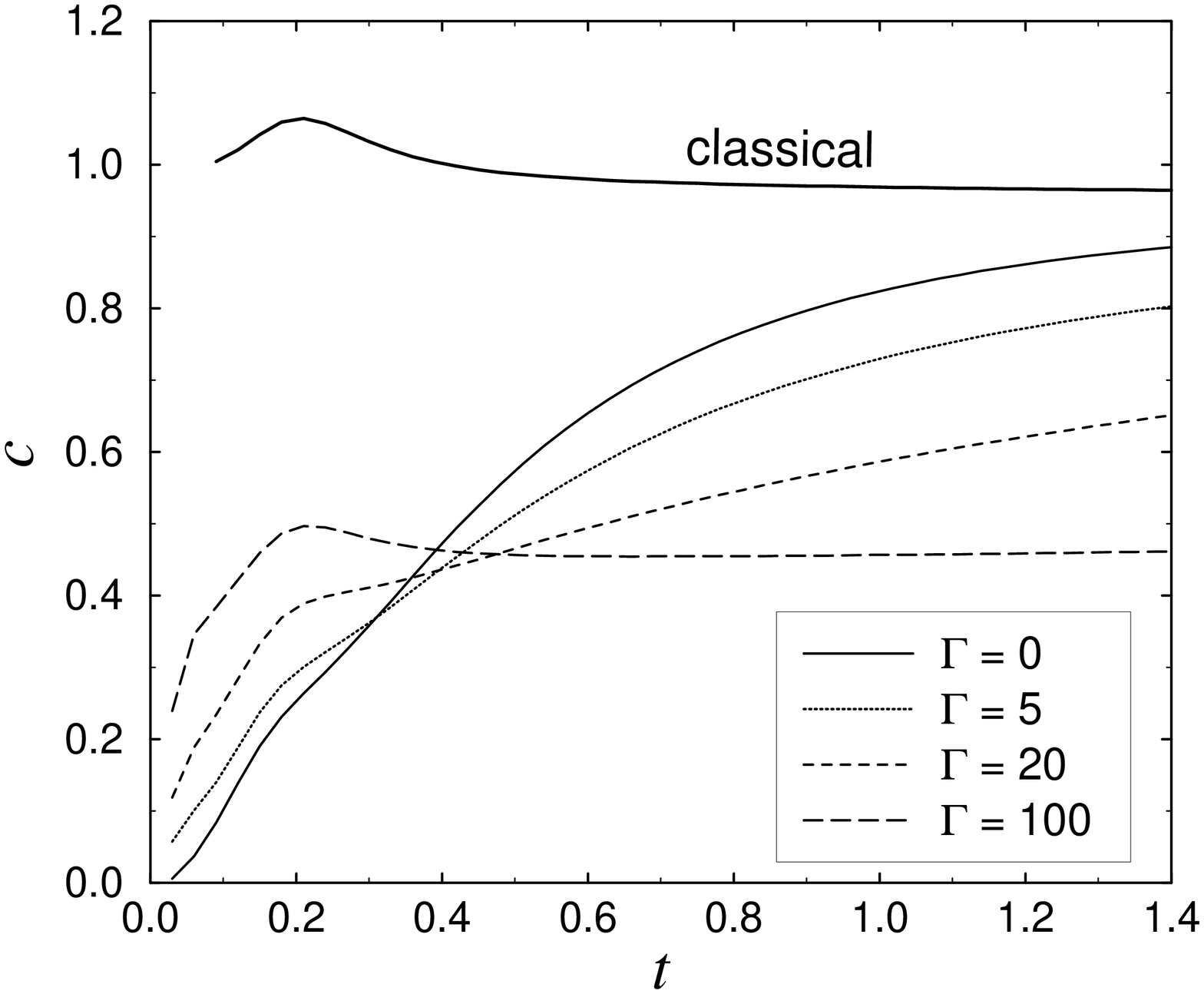,width=59mm,angle=270}
\hfill
\psfig{bbllx=20mm,bblly=20mm,bburx=188mm,bbury=238mm,%
figure=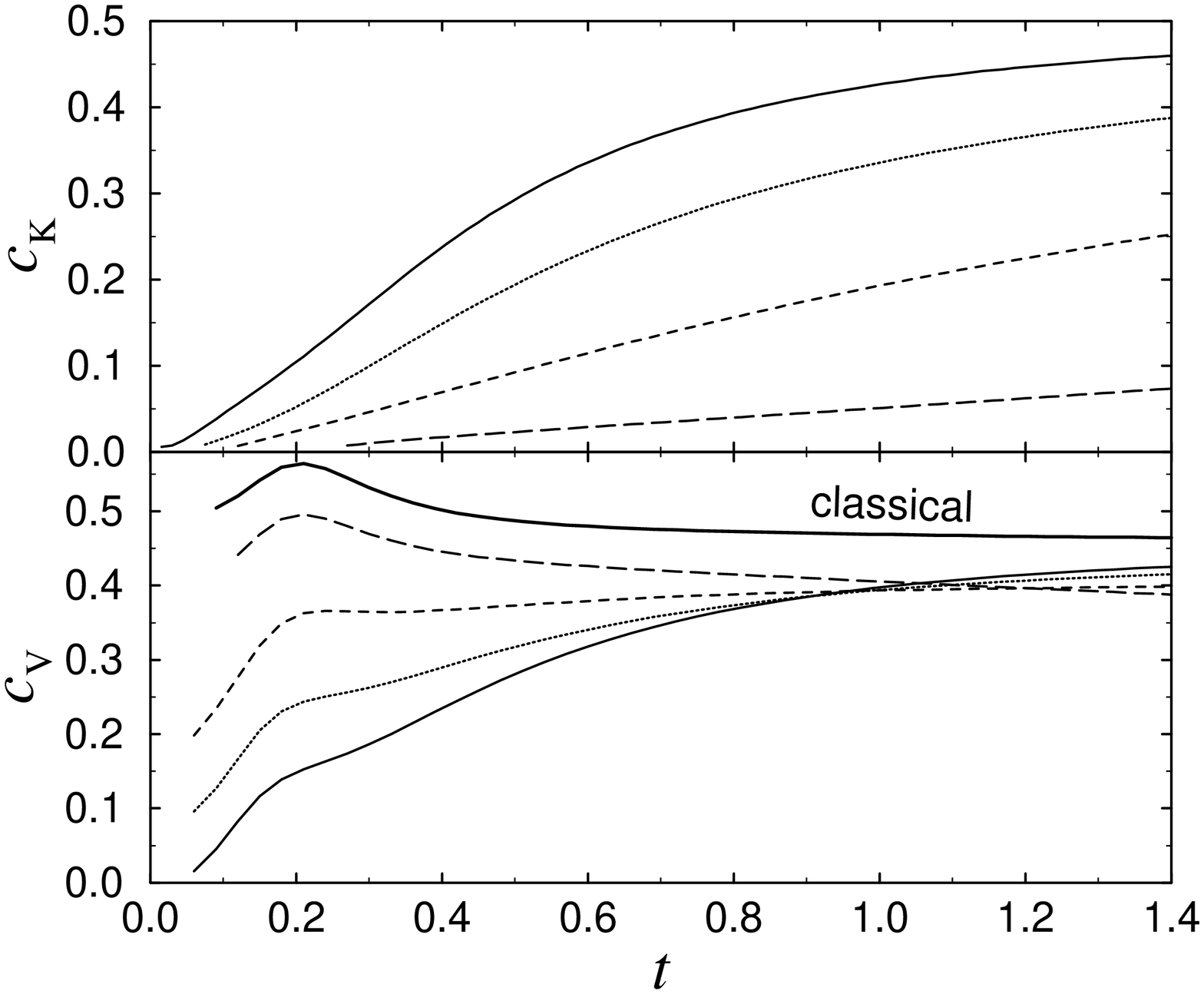,width=59mm,angle=270}}
\caption{Total specific heat $c(t)$ and the kinetic and interaction
parts of the specific heat, namely
$c_{{}_{\rm{K}}}(t)=\partial_t\langle\hat{K}\rangle$ and
$c_{{}_{\rm{V}}}(t)=\partial_t\langle\hat{V}\rangle$ vs. reduced
temperature $t$, for different values of the damping strength
$\Gamma=\gamma/\Omega$. Note that for $\Gamma\to\infty$, $c(t)$ tends
to $c_{\rm{cl}}-1/2$: this behavior can be explained by observing
that, in the strong damping limit,
$c_{{}_{\rm{K}}}(t)\to{0}$.\label{f.2}}
\end{minipage}
\end{figure}

Again, we observe that the fluctuations of
coordinate-dependent observables are quenched by increasing the
dissipation strength $\gamma$, while those of momentum-dependent ones
are enhanced due to the momentum exchanges with the environment. The
result of these opposite behaviors is a non-trivial dependence on
dissipation of ``mixed'' quantities like the specific heat as is shown
in Fig.~\ref{f.2}\,.
\medskip

The last system we consider is the dissipative quantum XY
model. Much interest in such system is due to its close relation with
2D granular superconductors and Josephson junction arrays (JJA). In the
last 15 years much attention has been devoted to the theoretical
and experimental study~\cite{jja} of the phase ordering in JJA
and of how it is influenced by the environmental coupling.
The undamped system is described by the action
\begin{equation}
 S_{\rm JJA}=\int_0^\beta du \bigg\{\frac12 \sum_{ij}{\dot \phi_i(u)
 \frac{C_{ij}}{4 e^2} \dot \phi_j(u)} - J \sum_{<ij>}
 \cos[\phi_i(u)-\phi_j(u)] \bigg\} ~,
\label{e.SJJA}
\end{equation}
where $\phi_i$ is the superconducting phase of the $i$-th island,
${<}ij{>}$ restricts the sum over nearest-neighbor bonds, $J$ is the
Josephson coupling,
$C_{ij}=(C_0+4C_1)\delta_{ij}-C_1\delta_{ij}^{({\rm{nn}})}$,
$C_0$ and $C_1$ are the self and mutual capacitance between
superconducting islands and $\delta_{ij}^{({\rm{nn}})}=1$ for
nearest neighbors, zero otherwise.

From Eq.~(\ref{e.SJJA}) two contributions to the energy in this system
can be observed: the kinetic one, due to the charging energy between
islands ($E_{\rm{c}} =q^2/2C$, $q=2e$), and the potential energy, due
to the Josephson coupling in the superconducting junctions. When
$E_{\rm{c}}{\ll}J$, the charges on each island fluctuate independently
from the phases $\phi_i$\,: the latter have then a classical XY behavior
and the associated Berezinskii-Kosterlitz-Thouless (BKT) phase transition
takes place at temperature $T^{\rm (cl)}_{{}_{\rm{BKT}}}\sim{0.892}\,J$.
In the opposite limit, the energy cost to transfer charges between
neighboring islands is too high, so the charges tend to be localized
and the phase ordering tends to be suppressed.

\begin{figure}[b!]
\begin{minipage}[t]{119mm}
\centerline{\psfig{bbllx=45mm,bblly=26mm,bburx=188mm,bbury=238mm,%
figure=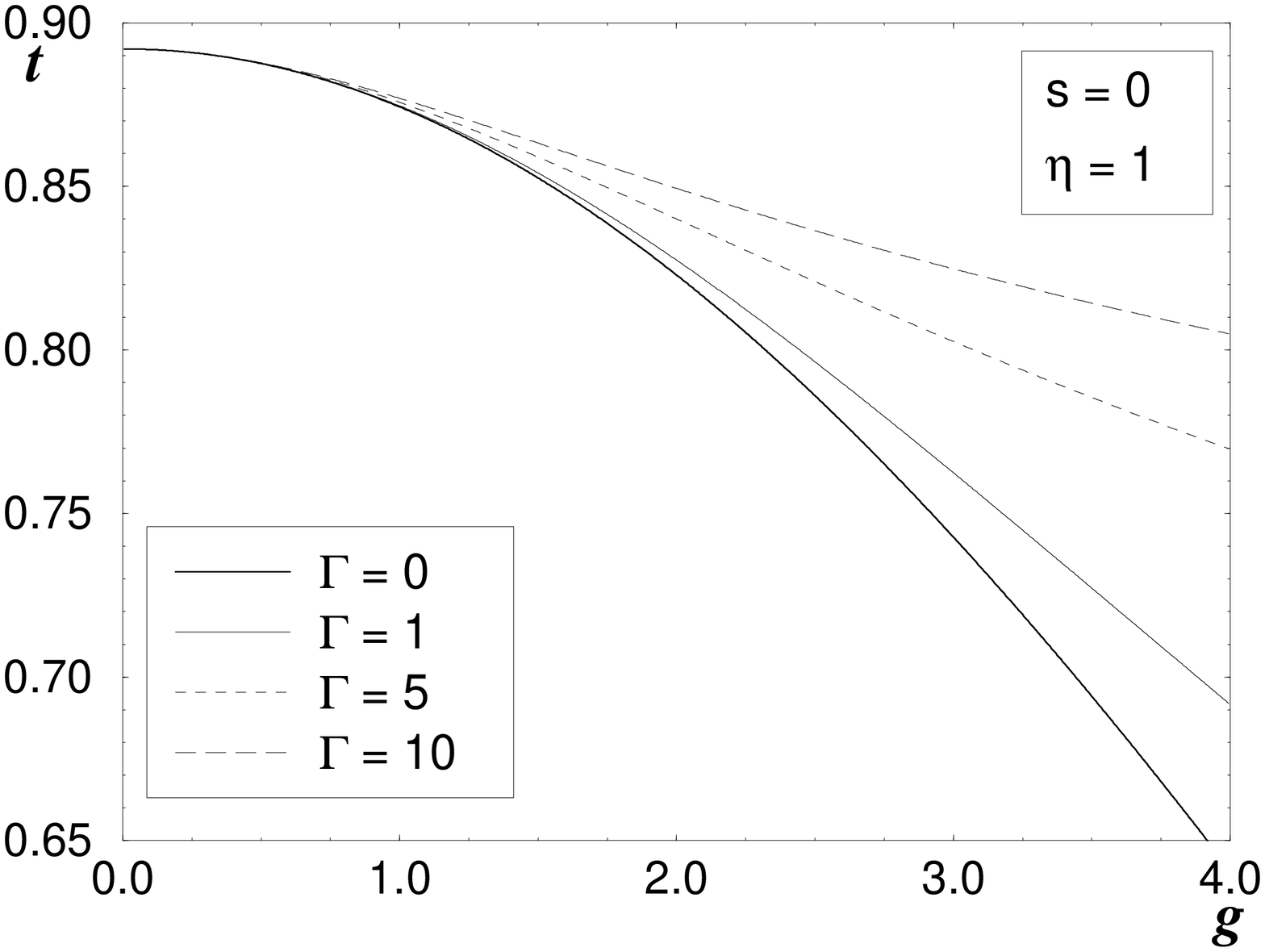,width=59mm,angle=270}
\hfill
\psfig{bbllx=45mm,bblly=26mm,bburx=188mm,bbury=238mm,%
figure=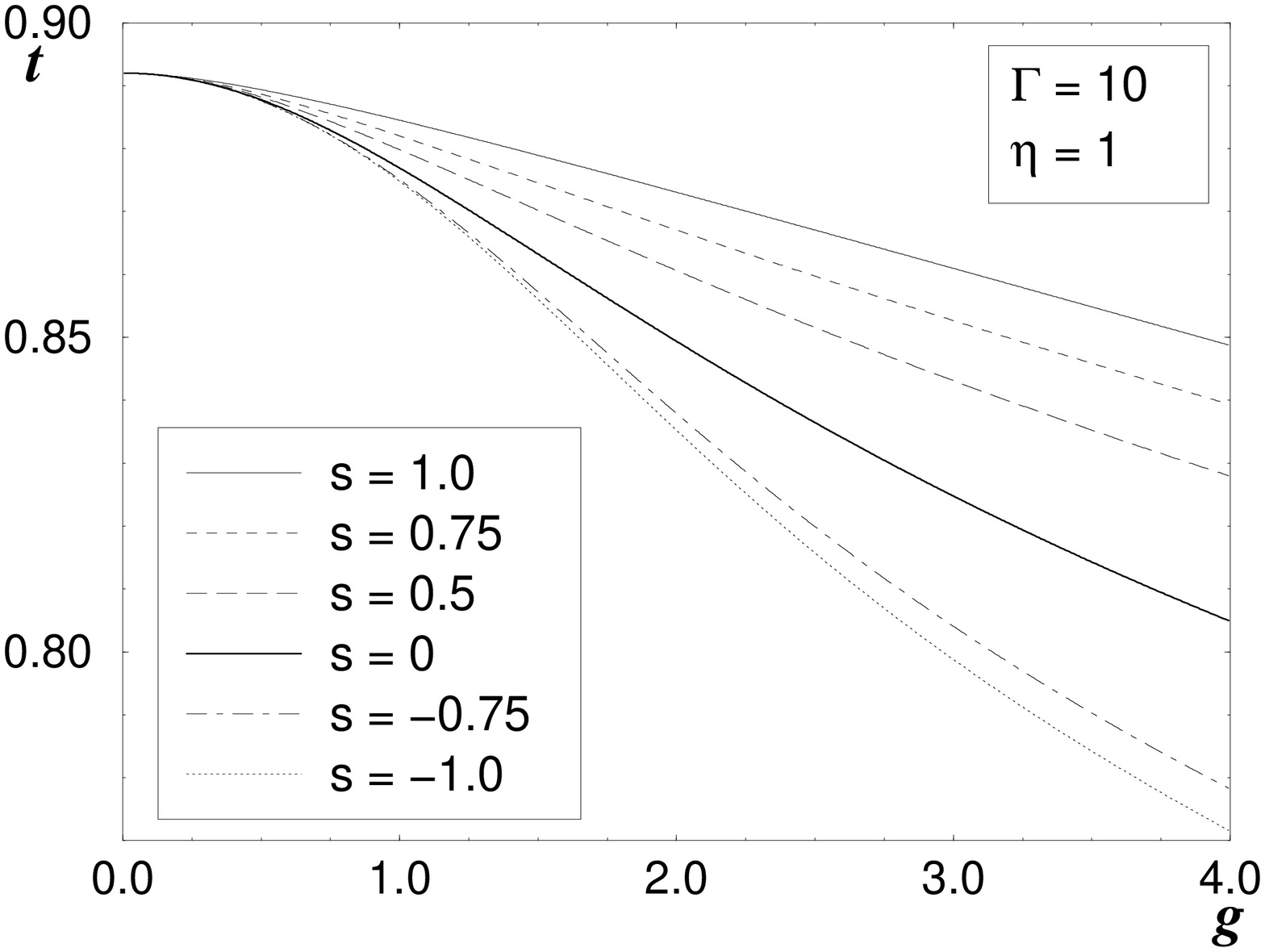,width=59mm,angle=270}}
\caption{Phase diagram in the $(g,t)$ plane, $t\equiv{T/J}$, for
different values of the damping parameters, $\Gamma=\gamma/\omega_0$
and $s$. On the left, the case of Ohmic dissipation: the critical temperature
$T_{{}_{\rm{BKT}}}$ tends to $T^{\rm (cl)}_{{}_{\rm{BKT}}}$ for increasing
damping strength. On the right, for different values of $s$: the
cases $s=1$ and $s=-1$ are nondissipative, since they correspond to
a variation of the capacitance- and of the frequency spectrum, respectively.
\label{f.3}}
\end{minipage}
\end{figure}

In this scenario dissipation is supposed to have an important
role. The environmental interaction tends to suppress the quantum
fluctuations~\cite{shon/tink} of $\phi_i$ and to restore an almost
classical BKT phase transition. Nevertheless, it is not clear which is
the physical mechanism of the dissipation.  For the single Josephson
junction with Ohmic dissipation the classical ``resistively and
capacitively shunted junction'' (RCSJ) model is
recovered~\cite{shon/tink}, but in the case of many degrees of freedom
the environmental interaction is much more complicated, e.g.
non-exponential memory effects and then non-Ohmic damping can
appear.
The dissipation model we assume consists in independent
environmental baths, one for each junction (or bond).
The dissipative part of the action is
\begin{equation}
 S_{\rm D}={\textstyle \frac12}\, {\sum}_{n}\, {^t\boldsymbol{\phi}_n} \boldsymbol{K}_n
 \boldsymbol{\phi}_n \,,
\label{e.SD}
\end{equation}
where the Fourier transform of the CL kernel matrix is given by
\begin{equation}
K_{n,ij} = \frac{\gamma}{2\pi}\,
\bigg(\frac{|\nu_n|}{\omega_0}\bigg)^{1+s}
(4~\delta_{ij} - \delta_{ij}^{\rm (nn)}) ~,
\label{e.Kn}
\end{equation}
and $\omega_0$ is a characteristic frequency, that we
choose as the Debye frequency
\begin{equation}
 \omega_0 \equiv \omega_{\pi,\pi}
 = 4 \bigg(\frac{q^2}{2 C_0} \frac{J}{1+8\eta}\bigg)^{1/2}~,
 ~~~~~\eta\equiv\frac{C_1}{C_0}.
\end{equation}
Now it is possible to define the quantum coupling parameter, that
measures the ``quanticity'' of the system as the ratio
between the characteristic quantum and classical energy scales
$g=\hbar\omega_0/J$.  The effective potential is calculated
using the extension to the many-body case~\cite{CFTV99} of the
scheme of Sec.~\ref{PQSCHA} and, apart from uniform terms, is given by
\begin{equation}
 V_{\rm eff}=-J_{\rm{eff}}(T) \sum_{<ij>} \cos(\phi_i-\phi_j) ~,
\label{e.Veff}
\end{equation}
where $J_{\rm{eff}}(T)=J\,e^{-D_1(T)/2}$ and $D_1(T)$ is the renormalization
parameter that measures the pure-quantum contribution to the fluctuations
of the nn relative superconducting phases.
The phase diagram of the system is calculated starting from the classical
effective potential (\ref{e.Veff}), i.e. by solving $T_{{}_{\rm{BKT}}}
=J_{\rm{eff}}(T_{{}_{\rm{BKT}}})\,T^{\rm (cl)}_{{}_{\rm{BKT}}}$\,,
and is shown in Fig.~\ref{f.3}, for different values of the
parameters that characterize the dissipation.

\newpage

\end{document}